\documentclass[a4paper]{article}
\usepackage{amssymb}

\usepackage{graphicx}
\usepackage{amsmath}
\usepackage{doublespace}
\usepackage{a4}

\input{tcilatex}

\begin{document}

\bigskip

\bigskip

\bigskip

\bigskip

\bigskip

\bigskip

\bigskip

\bigskip

\begin{center}
\ \ \ \ \ \ \ \ \ \ \ \ \ \ {\large Relativistic Particles and the
Cosmological Constant}\ {\large \bigskip }

\bigskip\ \ \ \ \ \ \ \ W. F. Chagas-Filho$^{\dagger }$

\bigskip\ \ \ \ \ \ \ \ \ \ \ \ Departamento de Fisica, Universidade Federal
de Sergipe, SE, Brazil
\end{center}

\ \ \ \ \ \ \ \ \ \ \ \ \ \ \ \ \ \ \ \ \ Abstract:\ \ We study a general
relativistic particle action obtained by incorporating the Hamiltonian
constraints into the formalism as a toy model for general relativity and
string theory. We show how a non-vanishing cosmological constant and a
weakening of gravity at short distances may be interpreted as evidences for
the existence of gravitational dipoles.

\bigskip

\bigskip

\bigskip

\bigskip

$\bigskip $

$\bigskip $

$\bigskip $

$\bigskip $

$\bigskip $

$\dagger $E-mail: wfilho@fisica.ufs.br


\bigskip

\bigskip

\bigskip

\bigskip

\section{Introduction}

In a recent paper, A. Zee [1] presents an interesting discussion about the
physical implications of a measured [2] [3] non-vanishing value for the
cosmological constant. Besides the possibility of the existence of the
gravitipole [4], the gravitational analogue of Dirac's magnetic monopole,
the author in [1] also considers the possibility that Einstein's tensor
gravitational field $g_{\mu \nu }$ may not be a fundamental field. According
to [1], there is an open possibility that the graviton is a particle
composed of more fundamental objects, and these fundamental objects may also
be '' as observable or as unobservable as the quarks''.

The idea of a composed graviton, as a consequence of a non-vanishing
cosmological constant, is consistent with a calculation presented by Siegel
[5]. It is shown in [5] that there exist stringy corrections to quantum
gravity at very short distances. These corrections would also be present [5]
in a composed gravity. According to [5], the corrections are so powerful
that only a scalar ''graviton '' need to be considered. The same result is
obtained by vertex techniques and by scattering amplitudes: the
gravitational force weakens at short distances [5]. One of the consequences
is that black-hole formation is avoided, and this agrees with a previous
work [6].

In this work we study relativistic particle theory as a toy model for the
gravitational and string theories. There are theoretical advantages in using
such a framework because the particle mass may be viewed as playing the role
of a \ ''cosmological constant '' [7]. The most common string formulation,
the Nambu-Goto formulation [8] 
\begin{equation*}
S=-T\int d\tau d\sigma \sqrt{(\dot{x}.\acute{x})^{2}-\dot{x}^{2}\acute{x}^{2}%
}
\end{equation*}
does not have a cosmological constant because it is simply the area of a
sheet. The alternative string action [9] 
\begin{equation*}
S=-\frac{T}{2}\int d\tau d\sigma \sqrt{-g}g^{ab}\partial _{a}x.\partial _{b}x
\end{equation*}
which is equivalent to describe the coupling of $D$ massless scalar fields $%
x^{\mu }$ , ( $\mu =0,1,...,D-1$ ) with gravity in $d=2$ dimensions, also
does not have a two-dimensional \ ''cosmological constant '' because it
breaks the Weyl invariance of $S$ and as a result leads to inconsistent
classical field equations. A cosmological constant appears in the
relativistic membrane action [10] 
\begin{equation*}
S=-\frac{T}{2}\int d\tau d^{2}\sigma (\sqrt{-g}g^{AB}\partial _{A}x.\partial
_{B}x-\Lambda \sqrt{g})
\end{equation*}
but membrane theory is a difficult theory and appears to be
non-renormalizable. Another advantage is that, in a second-quantized
formalism, relativistic particles are described by scalar fields. If we then
consider also massless particles, this will bring us very close to the
propagator results [5] for the scalar ''graviton''.

In this work we will also be interested in the relation between the
presence, or the absence, of a \ ''cosmological constant '' in the particle
action and the space-time invariances of the action. As we will see, when
the particle's mass is equal to zero, the action has a larger set of
space-time invariances. A non-zero mass value is then related to the
breaking of some, or all, of these extra invariances. The paper is organized
as follows. In section two we review the most general solution of the
Killing equation for a flat $D=4$ space time and the corresponding
space-time algebra. Section three and four deal with massive relativistic
particles while in sections five and six we consider massless particles. The
last section contains some conclusions.

\bigskip

\bigskip

\bigskip

\section{General space-time algebra}

Consider a Riemannian manifold with a metric of Euclidean signature 
\begin{equation}
ds^{2}=g_{\mu \nu }(x)dx^{\mu }dx^{\nu }  \tag{1}
\end{equation}
Under an infinitesimal coordinate transformation 
\begin{equation}
x^{\mu }\rightarrow x^{\mu }+\delta x^{\mu }  \tag{2}
\end{equation}
we have 
\begin{equation}
\delta g_{\mu \nu }=-(D_{\mu }\delta x_{\nu }+D_{\nu }\delta x_{\mu }) 
\tag{3}
\end{equation}
where 
\begin{equation}
D_{\mu }\delta x_{\nu }=\partial _{\mu }\delta x_{\nu }-\Gamma _{\mu \nu
}^{\lambda }\delta x_{\lambda }  \tag{4}
\end{equation}
and 
\begin{equation}
\Gamma _{\mu \nu }^{\lambda }=\frac{1}{2}g^{\lambda \delta }(\partial _{\mu
}g_{\delta \nu }+\partial _{\nu }g_{\mu \delta }-\partial _{\delta }g_{\mu
\nu )}  \tag{5}
\end{equation}

Let us next consider the vector field 
\begin{equation}
c(\epsilon )=\epsilon ^{\mu }(x)\partial _{\mu }  \tag{6}
\end{equation}
such that 
\begin{equation}
D_{\mu }\epsilon _{\nu }+D_{\nu }\epsilon _{\mu }=\frac{2}{d}g_{\mu \nu
}D.\epsilon  \tag{7}
\end{equation}
where $d$ is the number of space-time dimensions. The vector field $%
c(\epsilon )$ generates the coordinate transformation 
\begin{equation}
\delta x^{\mu }=c(\epsilon )x^{\mu }=\epsilon ^{\mu }(x)  \tag{8}
\end{equation}
such that 
\begin{eqnarray}
\delta g_{\mu \nu } &=&-(D_{\mu }\epsilon _{\nu }+D_{\nu }\epsilon _{\mu }) 
\notag \\
&=&-\frac{2}{d}D.\epsilon (x)g_{\mu \nu }=\alpha (x)g_{\mu \nu }  \TCItag{9}
\end{eqnarray}
The effect of such a transformation is to scale the length of each vector
and it can be shown that it preserves the angle between the vectors. This
transformation is known as a conformal (angle-preserving) transformation.

Let us now restrict ourselves to the $d=4$ flat space-time. In this case
equation (7) becomes 
\begin{equation}
\partial _{\mu }\epsilon _{\nu }+\partial _{\nu }\epsilon _{\mu }=\frac{2}{d}%
\delta _{\mu \nu }\partial .\epsilon  \tag{10}
\end{equation}
One can show that the most general solution for this equation is [11] 
\begin{equation}
\epsilon ^{\mu }=a^{\mu }+\omega ^{\mu \nu }x_{\nu }+\alpha x^{\mu
}+(2x^{\mu }x^{\nu }-\delta ^{\mu \nu }x^{2})b_{\nu }  \tag{11}
\end{equation}
so that 
\begin{equation}
c(\epsilon )=a^{\mu }P_{\mu }-\frac{1}{2}\omega ^{\mu \nu }M_{\mu \nu
}+\alpha D+b^{\mu }K_{\mu }  \tag{12}
\end{equation}
where 
\begin{eqnarray}
P_{\mu } &=&\partial _{\mu }  \TCItag{13} \\
M_{\mu \nu } &=&(x_{\mu }\partial _{\nu }-x_{\nu }\partial _{\mu }) 
\TCItag{14} \\
D &=&x^{\mu }\partial _{\mu }  \TCItag{15} \\
K_{\mu } &=&(2x_{\mu }x^{\nu }-\delta _{\mu }^{\nu }x^{2})\partial _{\nu } 
\TCItag{16}
\end{eqnarray}
$P_{\mu }$ generates space-time translations, $M_{\mu \nu }$ is the
generator of four-dimensional rotations, $D$ is the generator of dilatations
and $K_{\mu }$ generates conformal transformations ( conformal boosts ).

These generators obey the commutator algebra 
\begin{eqnarray}
\lbrack P_{\mu },P_{\nu }] &=&0  \TCItag{17} \\
\lbrack P_{\mu },M_{\nu \lambda }] &=&\delta _{\mu \nu }P_{\lambda }-\delta
_{\mu \lambda }P_{\nu }  \TCItag{18} \\
\lbrack M_{\mu \nu },M_{\lambda \rho }] &=&\delta _{\nu \lambda }M_{\mu \rho
}+\delta _{\mu \rho }M_{\nu \lambda }  \TCItag{19} \\
\lbrack D,D] &=&0  \TCItag{20} \\
\lbrack D,P_{\mu }] &=&-P_{\mu }  \TCItag{21} \\
\lbrack D,M_{\mu \nu }] &=&0  \TCItag{22} \\
\lbrack D,K_{\mu }] &=&K_{\mu }  \TCItag{23} \\
\lbrack P_{\mu },K_{\nu }] &=&2(\delta _{\mu \nu }D-M_{\mu \nu }) 
\TCItag{24} \\
\lbrack M_{\mu \nu },K_{\lambda }] &=&\delta _{\lambda \nu }K_{\mu }-\delta
_{\lambda \mu }K_{\nu }  \TCItag{25} \\
\lbrack K_{\mu },K_{\nu }] &=&0  \TCItag{26}
\end{eqnarray}
This is the full $d=4$ Euclidean flat space-time algebra. Later we will show
that a new invariance of the free massless particle allows a
four-velocity-dependent extension for this space-time algebra.

\section{Relativistic particles}

A relativistic particle describes in space-time a one-parameter trajectory $%
x^{\mu }(\tau )$. The action must be independent of the parameter choice and
so is taken to be proportional to the arc-length traveled by the particle.
For a relativistic particle in a flat space-time this action is given by 
\begin{equation}
S=-m\int ds=-m\int d\tau \sqrt{-\dot{x}^{2}(\tau )}  \tag{27}
\end{equation}
where $\tau $ is a parameter, $m$ is the particle's mass, $ds^{2}=-\delta
_{\mu \nu }dx^{\mu }dx^{\nu }$ and $\dot{x}^{\mu }=\frac{dx^{\mu }}{d\tau }$%
. We use units in which $\hbar =c=1$ .

Action (27) is invariant under Poincar\'{e} transformations 
\begin{equation}
\delta x^{\mu }=a^{\mu }+\omega _{\nu }^{\mu }x^{\nu }  \tag{28}
\end{equation}
with $\omega ^{\mu \nu }=-\omega ^{\nu \mu }$ , and under reparametrizations
of the world-line 
\begin{equation}
\tau \rightarrow \tau ^{\prime }=f(\tau )  \tag{29}
\end{equation}
where $f$ \ is an arbitrary function. Action (27) defines a non-linear,
one-dimensional, field theory with many subtleties Requiring a stationary
action under $x^{\mu }$ variations is equivalent to requiring the vanishing
of the symmetric tensor 
\begin{equation}
\sigma _{\mu \nu }=\ddot{x}_{\mu }\dot{x}_{\nu }+\ddot{x}_{\nu }\dot{x}_{\mu
}  \tag{30}
\end{equation}
The trivial solution is a free relativistic particle with 
\begin{equation}
\ddot{x}^{\mu }=0  \tag{31}
\end{equation}
but, in principle, non-trivial solutions may also exist.

Action (27) is obviously inadequate if we wish to study the massless limit
of relativistic particle theory.\ We can formally construct a more general
action, compatible with the $m=0$ limit, by treating the relativistic
particle as a constrained Hamiltonian system.

In the transition to the Hamiltonian formalism action (27) gives the
canonical momentum 
\begin{equation}
p_{\mu }=\frac{m}{\sqrt{-\dot{x}^{2}}}\dot{x}_{\mu }  \tag{32}
\end{equation}
and this momentum gives rise to the primary Hamiltonian constraint [12] 
\begin{equation}
\phi =\frac{1}{2}(p^{2}+m^{2})\approx 0  \tag{33}
\end{equation}
In this work we follow Dirac's [12] convention that a constraint is set
equal to zero only after all calculations have been performed. Equation (33)
means that $\phi $ ''weakly '' vanishes.

The canonical Hamiltonian corresponding to the Lagrangian of action (27), $%
H=p.\dot{x}-L$, identically vanishes. This is due to the reparametrization
invariance of action (27). Dirac's extended Hamiltonian is then 
\begin{equation}
H_{E}=H+\lambda \phi =\frac{\lambda }{2}(p^{2}+m^{2})  \tag{34}
\end{equation}
where $\lambda (\tau )$ is a Lagrange multiplier enforcing the constraint $%
\phi $. We can then define the following Lagrangian for the relativistic
particle 
\begin{eqnarray}
L &=&p.\dot{x}-H_{E}  \notag \\
&=&p.\dot{x}-\frac{1}{2}\lambda (p^{2}+m^{2})  \TCItag{35}
\end{eqnarray}
Lagrangian (35) is defined in an extended configuration space containing
also $p_{\mu }$ as a generalized coordinate. To eliminate $p^{\mu }$ we
construct a variational principle based on (35). The solution of the
equation of motion is 
\begin{equation}
p_{\mu }=\frac{1}{\lambda }\dot{x}_{\mu }  \tag{36}
\end{equation}
Inserting this result into (35) we obtain the relativistic particle action 
\begin{equation}
S=\frac{1}{2}\int d\tau (\lambda ^{-1}\dot{x}^{2}-\lambda m^{2})  \tag{37}
\end{equation}
Action (37) formally coincides with the ``einbein'' action 
\begin{equation*}
S=\frac{1}{2}\int d\tau (e^{-1}\dot{x}^{2}-em^{2})
\end{equation*}
in which the auxiliary variable $e(\tau )$ describes [13] the
one-dimensional geometry of the particle's world-line and $m^{2}$ plays the
role of a \ \ ''cosmological constant '' [7]. One way to investigate the
role played by the one-dimensional field $\lambda (\tau )$ in action (37) is
to construct the corresponding Hamiltonian formalism. Action (37) gives a
canonical momentum identical to (36) and the primary constraint 
\begin{equation}
p_{\lambda }=\frac{\delta S}{\delta \dot{\lambda}}=0  \tag{38}
\end{equation}
The canonical Hamiltonian that corresponds to the Lagrangian of action (37)
is 
\begin{equation}
H=\frac{1}{2}\lambda (p^{2}+m^{2})  \tag{39}
\end{equation}
Constraint (38) must now be incorporated into the formalism. We then get the
extended Hamiltonian 
\begin{eqnarray}
H_{E} &=&H+\zeta p_{\lambda }  \notag \\
&=&\frac{1}{2}\lambda (p^{2}+m^{2})+\zeta p_{\lambda }  \TCItag{40}
\end{eqnarray}
where $\zeta $ is a Lagrange multiplier. We must now develop Dirac's
algorithm for constrained systems, which means to require the stability of
all possible constraints. The stability of constraint (38), $\dot{p}%
_{\lambda }=\{p_{\lambda },H_{E}\}=0$ is satisfied only if $\phi =\frac{1}{2}%
(p^{2}+m^{2})\approx 0$ . $\phi $ is then a secondary constraint for action
(37). Since $\phi $ already appears in $H_{E}$ , it is not necessary to
incorporate it into the formalism, and its stability condition is simply $%
\dot{\phi}=\{\phi ,H_{E}\}=0$. This condition is automatically satisfied
because $\phi $ has vanishing Poisson bracket with itself and with $%
p_{\lambda }=0$ ( $\phi $ is a first-class constraint ). The algorithm is
then completed and no condition is placed on $\lambda (\tau )$. It remains
as an arbitrary function in action (37). Constraint (38), which is also
first-class, generates arbitrary translations of $\lambda $. If we count the
physical degrees of freedom according to the method proposed in [14], we
will find that there are no physical degrees of freedom associated to the
canonical pair $(\lambda ,p_{\lambda })$ and that only $D-1$ of the $D$ $%
x^{\mu }$'s are physical.

If we vary $\lambda $ in action (33) we obtain the relation 
\begin{equation}
\frac{\dot{x}^{2}}{\lambda ^{2}}+m^{2}=0  \tag{41}
\end{equation}
which, in view of equation (36), reproduces constraint $\phi $ of equation
(33). From equation (41) we get 
\begin{equation}
\lambda =\frac{\sqrt{-\dot{x}^{2}}}{m}  \tag{42}
\end{equation}
and inserting this result back into action (37), it becomes identical to
action (27). Actions (37) and (27) are therefore equivalent at the classical
level. However, action (37) is more general because it allows a transition
to the $m=0$ limit of the theory.

Action (37) is invariant under the Poincar\`{e} transformation 
\begin{eqnarray}
\delta x^{\mu } &=&a^{\mu }+\omega _{\nu }^{\mu }x^{\nu }  \TCItag{43a} \\
\delta \lambda &=&0  \TCItag{43b}
\end{eqnarray}
and under the infinitesimal reparametrizations 
\begin{eqnarray}
\delta x^{\mu } &=&\epsilon \dot{x}^{\mu }  \TCItag{44a} \\
\delta \lambda &=&\frac{d}{d\tau }(\epsilon \lambda )  \TCItag{44b}
\end{eqnarray}
where $\epsilon (\tau )$ is an arbitrary parameter. The presence of a
definite mass value spoils the invariance of the relativistic particle
action (37) under space-time scale and conformal transformations. The
massive particle action (37) is then invariant only under a sub-algebra of
the full space-time algebra defined by equations (17) to (26). This
sub-algebra is given by 
\begin{eqnarray}
\lbrack P_{\mu },P_{\nu }] &=&0  \TCItag{45a} \\
\lbrack P_{\mu },M_{\nu \lambda }] &=&\delta _{\mu \nu }P_{\lambda }-\delta
_{\mu \lambda }P_{\nu }  \TCItag{45b} \\
\lbrack M_{\mu \nu },M_{\lambda \varrho }] &=&\delta _{\nu \lambda }M_{\mu
\varrho }+\delta _{\mu \varrho }M_{\nu \lambda }  \TCItag{45c}
\end{eqnarray}

The Lagrangian equation of motion for $x^{\mu }$ that follows from action
(37) state that 
\begin{equation}
\frac{d}{d\tau }(\frac{1}{\lambda }\dot{x}^{\mu })=0  \tag{46}
\end{equation}
and we see that the free sector of the relativistic particle theory based in
action (37) corresponds to the subspace of configuration space where $\dot{%
\lambda}=0.$

It is interesting to examine the consequences, at the classical level, of
the fact that $\lambda $ must be a constant in the free theory. To do this,
we must go deep in the region of short distances and of relativistic
effects. For instance, from relation (42) we get 
\begin{equation}
-\lambda m^{2}=\lambda ^{-1}\dot{x}^{2}  \tag{47}
\end{equation}
so that the Lagrangian of action (37) becomes 
\begin{equation}
L=-\lambda m^{2}  \tag{48}
\end{equation}
and in the free sector this must be a constant. We can then take the
particle mass $m$ to be a multiple of a fundamental mass $\mu $ that
characterizes the energy region. The value $\lambda =0$ is ruled out by
equation (46). If we choose $\lambda =n^{2}$ , with $n=n_{+}+n_{-}$ and $%
n_{+}=1,2,3,...$and $n_{-}=-1,-2,-3,...$ we get the condition 
\begin{equation*}
(n_{+}+n_{-})\mu =cons\tan t
\end{equation*}
and with the particular choice $n_{+}=1$ , $n_{-}=-1$ we may construct a
massless scalar particle as composed of the sum of a more fundamental
particle of mass $\mu $ and a more fundamental particle of mass $-\mu $. At
very high energies the massless particle is a very small gravitational
dipole. The reader should recall that negative masses are allowed by the
relativistic equation ( in our system of unities ) 
\begin{equation*}
E^{2}=p^{2}+m^{2}
\end{equation*}

According to Newton's equation for the gravitational interaction, masses of
opposite signs should repel each other. The gravitational dipole can only be
stable under the influence of stronger forces, as protons are held together
by the strong nuclear force, despite their mutual electromagnetic repulsion.
The repulsion energy of the gravitational dipole will give rise to a very
tiny, but nonvanishing mass. We may then suspect that a non-vanishing
cosmological constant may be associated with repulsion energy, and that this
repulsion is the origin of the observed expansion of the universe.

\section{Quantization}

If we start a quantization process by elevating $x^{\mu }$ and $p_{\mu }$ to
operators 
\begin{eqnarray}
x^{\mu } &\rightarrow &x^{\mu }  \TCItag{47a} \\
p_{\mu } &\rightarrow &i\frac{\partial }{\partial x^{\mu }}  \TCItag{47b}
\end{eqnarray}
which satisfy the commutator relation 
\begin{equation}
\lbrack x^{\mu },p_{\nu }]=\delta _{\nu }^{\mu }  \tag{48}
\end{equation}
we must take care of the gauge invariance associated to the presence of the
first-class constraint $\phi $ . Following the Gupta-Bleuler quantization
method, we elevate $\phi $ to an operator $\phi _{op}$ using equations (47)
and require that this operator annihilates physical states 
\begin{equation}
\phi _{op}\mid phys\rangle =(\square -m^{2})\mid phys\rangle =0  \tag{49}
\end{equation}
Equation (49) is just the Klein-Gordon equation for the physical states.
This equation can be derived from the second-quantized Lagrangian 
\begin{equation}
L=\varphi (\square -m^{2})\varphi  \tag{50}
\end{equation}
where $\varphi =\langle \vec{x},t\mid phys\rangle $. Now, to get a
consistent quantum theory based on the classical action (37), we must also
take care of the gauge invariance associated with the constraint $p_{\lambda
}=0.$ We do this by making the gauge choice 
\begin{equation}
\lambda =1  \tag{51}
\end{equation}
The Green function for the quantum propagation of the relativistic particle
can then be written 
\begin{eqnarray}
\triangle _{F}(x_{1},x_{2}) &=&\langle x_{1}\mid \frac{1}{\square -m^{2}}%
\mid x_{2}\rangle  \TCItag{52a} \\
&=&\langle x_{1}\mid \int_{0}^{\infty }d\tau e^{-\tau (\square -m^{2})}\mid
x_{2}\rangle  \TCItag{52b} \\
&=&\int_{0}^{\infty }d\tau \int_{x_{1}}^{x_{2}}Dx\exp \{-\frac{1}{2}%
\int_{0}^{\tau }d\bar{\tau}(\dot{x}^{2}-m^{2})\}  \TCItag{52c} \\
&=&\int_{0}^{\infty }d\tau \int_{x_{1}}^{x_{2}}DxD\lambda \delta (\lambda -1)
\notag \\
&&\exp \{-\frac{1}{2}\int_{0}^{\tau }d\bar{\tau}(\lambda ^{-1}\dot{x}%
^{2}-\lambda m^{2})\}  \TCItag{52d} \\
&=&\int_{0}^{\infty }d\tau \int_{x_{1}}^{x_{2}}DxDpD\lambda \delta (\lambda
-1)  \notag \\
&&\times \exp \{-\int_{0}^{\tau }d\bar{\tau}[p.\dot{x}-\frac{1}{2}\lambda
(p^{2}+m^{2})]\}  \TCItag{52e}
\end{eqnarray}
However, choosing $\lambda =1$ \ does not completely fix the
reparametrization in (44b) and the consequence is that there still exists
propagating unphysical states. To be more precise we must include a
Faddeev-Popov [15] factor. As can be seen from (44b), the Faddeev-Popov
determinant associated with the gauge choice $\lambda =1$ is the determinant
of the derivative, which can be written as 
\begin{eqnarray}
\triangle _{FP} &=&\det \mid \partial _{\tau }\mid  \notag \\
&=&\int D\theta D\bar{\theta}\exp \{i\int d\tau \bar{\theta}\partial _{\tau
}\theta  \TCItag{53}
\end{eqnarray}
if we introduce anticommuting Grassmann ghost variables $\theta $ \ and \ $%
\bar{\theta}$ . Inserting the above expression in equation (52e) and
performing the $\lambda $ integration, we have the functional integral 
\begin{equation}
\int DxDpD\theta D\bar{\theta}\exp \{-\int_{0}^{\tau }[p.\dot{x}-\frac{1}{2}%
(p^{2}+m^{2})-i\bar{\theta}\partial _{\tau }\theta ]\}  \tag{54}
\end{equation}
The effective action in the exponential above is invariant under the BRST
[16] transformation 
\begin{eqnarray*}
\delta x^{\mu } &=&i\epsilon \theta \dot{x}^{\mu } \\
\delta p_{\mu } &=&i\epsilon \theta \dot{p}^{\mu } \\
\delta \theta &=&i\epsilon \theta \dot{\theta} \\
\delta \bar{\theta} &=&i\epsilon \theta (\bar{\theta}\dot{)}+\frac{1}{2}%
\epsilon (p^{2}+m^{2})
\end{eqnarray*}
and we see that the presence of the BRST invariance for the relativistic
particle is related to the fact that $\lambda $ must be a constant in the
free theory. As we saw above, a constant $\lambda $ may be related to a
non-vanishing ''cosmological constant '' , and the existence of the free
particle sector is a declaration of the existence of inertial frames. These
concepts are mixed with the origin of the BRST invariance.

\section{Massless relativistic particles}

A massless relativistic particle may be described by the action 
\begin{equation}
S=\frac{1}{2}\int_{\tau _{i}}^{\tau _{f}}d\tau e^{-1}\dot{x}^{2}  \tag{55}
\end{equation}
which is the $m=0$ limit of action (37). The equation of motion for $x^{\mu
}(\tau )$ that follows from (55) is identical to equation (46) and $\lambda $
must again be a constant in the free sector.

The massless action (55) is again invariant under the Poincar\'{e}
transformation (43) and under the reparametrization (44). It is also
invariant under the scale transformation 
\begin{eqnarray}
\delta x^{\mu } &=&\alpha x^{\mu }  \TCItag{56a} \\
\delta \lambda &=&2\alpha \lambda  \TCItag{56b}
\end{eqnarray}
and under the conformal transformation 
\begin{eqnarray}
\delta x^{\mu } &=&(2x^{\mu }x^{\nu }-\eta ^{\mu \nu }x^{2})b_{\nu } 
\TCItag{57a} \\
\delta \lambda &=&4\lambda x.b  \TCItag{57b}
\end{eqnarray}
The massless action has then two additional space-time invariances, and the
full space-time algebra (17-26) is satisfied. We may then suspect that the
appearance of a non-vanishing mass value is associated to the breaking of
the above invariances.

In ref. [17] it was shown that in the free sector, where $\dot{\lambda}=0$ ,
the massless action (55) is also invariant under the off-shell
transformation 
\begin{eqnarray}
x^{\mu } &\rightarrow &\exp \{\frac{1}{3}\beta (\dot{x}^{2})\}x^{\mu } 
\TCItag{58a} \\
\lambda  &\rightarrow &\exp \{\frac{2}{3}\beta (\dot{x}^{2})\}\lambda  
\TCItag{58b}
\end{eqnarray}
where $\beta $ is an arbitrary function of $\dot{x}^{2}$ . In the present
work we point out that this \ ''velocity''-dependent scale transformation
allows us to extend the generator of dilatations $D$ , of equation (15), to
include also ''velocity''-dependent dilatations, 
\begin{equation}
D\rightarrow D^{\ast }=D+\beta (\dot{x}^{2})D  \tag{59a}
\end{equation}
In fact, because $\beta $ \ is a function of $\dot{x}^{\mu }$ only, we can
also introduce the extended operators 
\begin{eqnarray}
P_{\mu }^{\ast } &=&P_{\mu }+\beta P_{\mu }  \TCItag{59b} \\
M_{\mu \nu }^{\ast } &=&M_{\mu \nu }+\beta M_{\mu \nu }  \TCItag{59c} \\
K_{\mu }^{\ast } &=&K_{\mu }+\beta K_{\mu }  \TCItag{59d}
\end{eqnarray}
corresponding to the space-time operators (13), (14) and (16). These new
operators obey the algebra 
\begin{equation}
\lbrack P_{\mu }^{\ast },P_{\nu }^{\ast }]=0  \tag{60a}
\end{equation}
\begin{equation}
\lbrack P_{\mu }^{\ast },M_{\nu \lambda }^{\ast }]=(\delta _{\mu \nu
}P_{\lambda }^{\ast }-\delta _{\mu \lambda }P_{\nu }^{\ast })+\beta (\delta
_{\mu \nu }P_{\lambda }^{\ast }-\delta _{\mu \lambda }P_{\nu }^{\ast }) 
\tag{60b}
\end{equation}
\begin{equation}
\lbrack M_{\mu \nu }^{\ast },M_{\lambda \rho }^{\ast }]=(\delta _{\nu
\lambda }M_{\mu \rho }^{\ast }+\delta _{\mu \rho }M_{\nu \lambda }^{\ast
})+\beta (\delta _{\nu \lambda }M_{\mu \rho }^{\ast }+\delta _{\mu \rho
}M_{\nu \lambda }^{\ast })  \tag{60c}
\end{equation}
\begin{equation}
\lbrack D^{\ast },D^{\ast }]=0  \tag{60d}
\end{equation}
\begin{equation}
\lbrack D^{\ast },P_{\mu }^{\ast }]=-P_{\mu }^{\ast }-\beta P_{\mu }^{\ast }
\tag{60e}
\end{equation}
\begin{equation}
\lbrack D^{\ast },M_{\mu \nu }^{\ast }]=0  \tag{60f}
\end{equation}
\begin{equation}
\lbrack D^{\ast },K_{\mu }^{\ast }]=K_{\mu }^{\ast }+\beta K_{\mu }^{\ast } 
\tag{60g}
\end{equation}
\begin{equation}
\lbrack P_{\mu }^{\ast },K_{\nu }^{\ast }]=2(\delta _{\mu \nu }D^{\ast
}-M_{\mu \nu }^{\ast })+2\beta (\delta _{\mu \nu }D^{\ast }-M_{\mu \nu
}^{\ast })  \tag{60h}
\end{equation}
\begin{equation}
\lbrack M_{\mu \nu ,}^{\ast }K_{\lambda }^{\ast }]=(\delta _{\lambda \nu
}K_{\mu }^{\ast }-\delta _{\lambda \mu }K_{\nu }^{\ast })+\beta (\delta
_{\lambda \nu }K_{\mu }^{\ast }-\delta _{\lambda \mu }K_{\nu }^{\ast }) 
\tag{60i}
\end{equation}
\begin{equation}
\lbrack K_{\mu }^{\ast },K_{\nu }^{\ast }]=0  \tag{60j}
\end{equation}
The new invariance (58) of the massless particle, which is present only when 
$\dot{\lambda}=0$ , allows the construction of an extension of the
space-time algebra (17-26). For a non-vanishing ''cosmological constant '' $m
$ , the above space-time algebra collapses to algebra (45).

\section{Quantization of massless particles}

The quantization process for massless particles may be carried out simply by
making $m=0$ in the corresponding equations for the massive particle. What
is important here is that because in a second-quantized formalism the
massless particle is described by the Lagrangian 
\begin{equation}
L=\varphi \square \varphi  \tag{61}
\end{equation}
we can reproduce here Siegel's results for the scalar ''graviton''. All we
have to do is to include a higher-derivative coupling with a source $T$ \
and write the total Lagrangian 
\begin{equation}
L=\varphi \square \varphi +\varphi \exp \{\frac{1}{4}\square \}T  \tag{62}
\end{equation}
Lagrangian (62) is equivalent to [5] 
\begin{equation}
L=\varphi \square \exp \{-\frac{1}{2}\square \}\varphi +\varphi T  \tag{63}
\end{equation}
and the gravitational potential is of the form [5] 
\begin{eqnarray}
\frac{1}{\square }\exp \{\frac{1}{2}\square \}\delta ^{3}(x) &\sim &\int
d^{3}pe^{ip.x}\int_{1}^{\infty }d\tau e^{-\tau p^{2}/2}  \TCItag{64} \\
&\sim &\int_{0}^{\infty }d\tau \tau ^{-3/2}e^{-r^{2}/2\tau }  \notag \\
&\sim &\frac{1}{r}\int_{0}^{r}d\sigma e^{-\sigma ^{2}/2}  \notag \\
&\sim &\frac{1}{r}\func{erf}(\frac{r}{\sqrt{2}})  \notag
\end{eqnarray}
In terms of the error function and where $\tau =r^{2}/\sigma ^{2}$ . The
result is that the attractive long-distance gravitational potential $-\frac{1%
}{r}$ \ is smoothed off to a parabola at short distances [5]. Could
repulsive gravitational forces be the responsible for this?

\section{Conclusion}

In this work we studied a general relativistic particle action, obtained by
incorporating the Hamiltonian constraints into the formalism, as a toy model
for the more complex gravitational and string theories. We showed that the
massless action has a larger set of space-time invariances and that in the
free massless sector an extended space-time algebra can be constructed. We
also discussed on the cosmological constant problem and its relations to
BRST invariance and to a weakening of gravity at very short distances.

\end{document}